# Beyond Experience Sampling: Evaluating Personal Informatics with Technology-Assisted Reconstruction

**Evangelos Karapanos**
Madeira Interactive Technologies Institute
Campus da Penteada
9020-105 Funchal, Portugal
E.Karapanos@gmail.com

## Abstract
Experience Sampling has been considered the golden standard of in-situ measurement, yet, at the expense of high burden to participants. In this paper we propose Technology-Assisted Reconstruction (TAR), a methodological approach that combines passive logging of users' behaviors with use of these data in assisting the reconstruction of behaviors and experiences. Through a number of recent and ongoing projects we will discuss how TAR may be employed for the evaluation of personal informatics systems, but also, conversely, how ideas from the field of personal informatics may contribute towards the development of new methodologies for in-situ evaluation.

## Introduction
The Experience Sampling Method is often referred to as the gold standard for the in-situ evaluation of personal informatics systems as it samples experiences and behaviors right at the moment of their occurrence, thus avoiding retrospection biases. However, it comes at a cost. It imposes high burden to participants and may also affect the actual experience (see [1] for a review of ES studies in the field of HCI).

Kahneman et al. [2] proposed the Day Reconstruction

| Reason | No |
|---|---|
| Disrupts the activity | 6 |
| Imposes high burden to participants | 3 |
| Requires high effort from researchers | 3 |
| Inappropriate for eliciting rich qualitative data | 3 |
| Misses rare and brief events | 3 |
| The user should be in control of when, what and how often to report | 2 |
| Limits sample size | 2 |
| Depends on participants' ability to articulate ongoing experience | 2 |
| Poses privacy concerns | 2 |
| Limits number of measured variables | 1 |
| Technology limitations | 1 |

**Table 1.** Reasons for not selecting the Experience Sampling Method (No of papers) [1].

Method (DRM) as a cost-effective alternative to ESM. DRM asks participants to list their daily activities as a continuous list of episodes. This provides a temporal context for reconstruction leading to an increased amount of contextual cues from which experiential information may be inferred. DRM has been found to provide a reasonably good approximation to experience sampling data, both in between and within-subject analyses, and the method has been well adopted in the HCI community.

With this position paper we argue that our field has the capacity to contribute towards a next step in the field of momentary assessment, that of *technology-assisted reconstruction*. This paradigm will combine passive logging of users' behaviors with use of these data in assisting the reconstruction process. We propose that we need to address two core questions: what *types of data* cue episodic memories as well as how *data representation* affects the reconstruction process. Our goal is to develop methods that eliminate any form of interruption during participants' whereabouts and daily experiences, while employing psychologically grounded processes for assisting the reconstruction of their behaviors and experiences.

### Technology-Assisted Reconstruction

Our work on technology-assisted reconstruction is motivated by a recent theory of how individuals recall emotions experienced in past events. This theory assumes that "emotional experience can neither be stored nor retrieved" [5, p. 935]. Instead, it assumes that people first retrieve contextual details from episodic memory and then infer emotions on the basis of this information (e.g., I recall myself screaming and having my hands raised while on the rollercoaster, thus I infer an experience of high arousal). It is thus suggested that through increasing the amount of recalled contextual cues from episodic memory, one could increase individuals' accuracy in recalling the exact emotions experienced during the event (see [3] for a more elaborate discussion on the topic).

Our first exploration of technology assisted reconstruction was through iScale, a survey tool that aimed at assisting users in reconstructing their past experiences with a product while positioning them along a timeline, thus using human memory as a source of longitudinal data on product adoption. In the development of iScale we employed two competing theoretical models of how people recall emotions and derived two different versions of iScale. Both approaches employed graphing as a means of guiding the reconstruction process over the past, however, each version imposed a distinct approach to reconstruction.

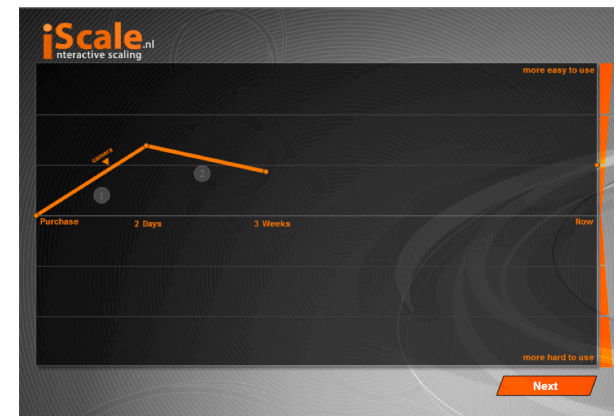

**Figure 1.** iScale.nl allows for reflection over long-term use with narrative data [3]. Two competing theories of emotion recall motivated the development of two versions of iScale, each imposing a particular process in reconstructing past experience.

For instance, the Constructive iScale which proved to be the most effective and was motivated by the theoretical approach we discussed earlier [5], asked participants to re-encounter their past experiences in a chronological sequence, starting from the moment of purchase of the product. Each time they would graph a segment, they would be asked to narrate one or more experiences that lead to this change in their perception. This forward chronological order as well as the concurrency of reporting and graphing was expected to lead to a richer recall from episodic memory, and consequently to a more accurate recall of emotions. The constructive iScale tool was found to result to more elicited experience reports compared to a control condition that employed no graphing, with more details (i.e., references to temporal information or particular events), but also to higher test-retest consistency in estimating when a particular event took place, as well as higher test-retest consistency on the graphed patterns over the other version of iScale (called Value-Account iScale).

Later, we applied the same principle in eliciting location-sharing preferences [4]. We hypothesized that trajectory reminders (i.e., locations visited before and after the location under study) would provide a temporal context to each recall, thus enriching the cues that one could recall from memory. 20 participants tracked their locations over four days. We extracted significant locations (i.e., ones they had spent more than 5 minutes within a 50 meter radius) and asked them to recall their location-sharing preferences across two conditions (with and without trajectory reminders, between subjects), employing varying levels of obfuscation (i.e., (1) do not share, (2) region, (3) city, (4) neighborhood, (5) exact address). After a week we asked participants to do the same task again. Overall, we found trajectory reminders to increase users' test-retest consistency in recalling location-sharing preferences.

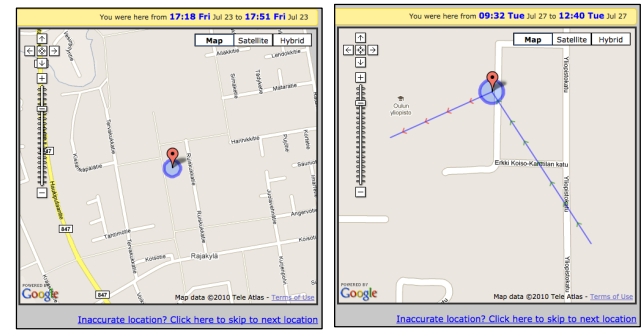

**Figure 2.** Providing temporal context with trajectory reminders (i.e., locations visited before and after) increases users' consistency in recall [4].

EmoSnaps (fig. 3) is a mobile application that captures self-face pictures unobtrusively and triggered by different events (e.g., slide in, respond or end call etc.). While several approaches to measuring emotion through facial expressions in mobile situations have been proposed, such as the one by Teeters, Kaliouby and Picard on "Self-Cam", a chest-mounted camera that is able to detect 24 feature points on the face and extract emotions using dynamic Bayesian Models, as well as Gruebler and Suzuki on a wearable interface device that can detect facial bioelectrical signals, with Emosnaps we aimed at a tool that is truly transparent to daily life and can be employed in longitudinal studies. In a two-week deployment we compared Emosnaps to a control condition (using only time information) while employing Experience Sampling as the ground truth. Emosnaps proved to result to a better approximation to Experience Sampling data than the control condition, however, surprisingly, participants were better able to accurately recall their emotions when using Emosnaps after one week and when the pictures were presented in random order than when reviewing the pictures at the end of the day in a chronological sequence. We hypothesized that this was due to a conflict between the process of inferring emotions from facial expressions and the one of inferring emotions from the recalled context of the day (as background cues in the photo would be better able to lead to result to a reconstruction of the context during the day but not after a week).

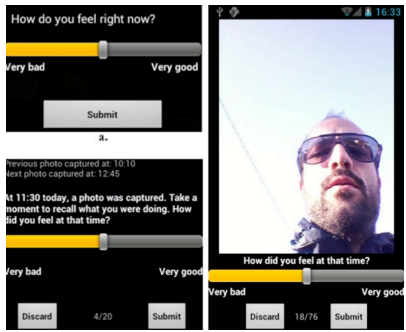

**Figure 3.** Emosnaps, a mobile application that captures unobtrusively users' self-face pictures throughout the day and uses them for the later recall of his or her emotions.

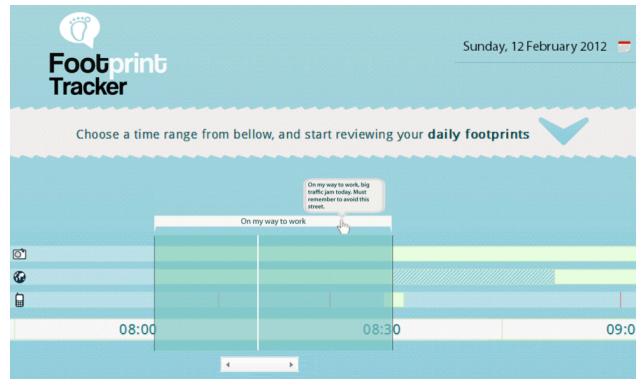

**Figure 4**. Footprint tracker, an online tool that supports users in reviewing logged data and reconstructing their past behaviors and experiences. It currently supports three types of data: visual (e.g., photos captured through Microsoft's Sensecam), location and context (i.e., Calls and SMSes made and received).

## Closing remarks

During our talk we will reflect on how technology-assisted reconstruction may be employed for the evaluation of personal informatics systems, but also, conversely, how ideas from the field of personal informatics may contribute towards the development of new methodologies for in-situ evaluation.